# Deducing Cardiorespiratory Motion of Cardiac Substructures Using a Novel 5D-MRI Workflow for Radiotherapy

**Chase Ruff, M.S.[1,2], Tarun Naren, M.S.[1], Oliver Wieben, Ph.D.[1,3], Prashant Nagpal, M.D.[3], Kevin Johnson, Ph.D.[1,3], Jiwei Zhao, Ph.D.[4,5], Thomas Grist M.D.[3], and Carri Glide-Hurst, Ph.D.[1,2]**

1) Department of Medical Physics, University of Wisconsin-Madison, Madison, Wisconsin
2) Department of Human Oncology, University of Wisconsin-Madison, Madison, Wisconsin
3) Department of Radiology, University of Wisconsin-Madison, Madison, Wisconsin
4) Department of Biostatistics and Medical Informatics, University of Wisconsin-Madison, Madison, Wisconsin
5) Department of Statistics, University of Wisconsin-Madison, Madison, Wisconsin
Email: Carri Glide-Hurst, Ph.D., glidehurst@humonc.wisc.edu

## Abstract

Objective: Cardiotoxicity is a devastating complication of thoracic radiotherapy, where dose to specific substructures is correlated with specific cardiovascular diseases. However, current practice ignores the radiosensitivities and complex motion trajectories of individual substructures. Additionally, current imaging protocols in radiotherapy are insufficient to decouple and quantify cardiac motion, limiting substructure-specific motion considerations in treatment planning. We propose a 5D-MRI workflow, capable of decoupling cardiorespiratory motion, for substructure-specific motion analysis, with future extension to margin calculation.

Approach: Our 5D-MRI workflow was implemented for 10 healthy volunteers, ranging from 23 to 65 years old, reconstructing images for end-exhale/inhale and active-exhale/inhale for end-systole/diastole. For motion assessment, proximal coronary arteries, chambers, and great vessels were contoured across all images and verified by a cardiovascular radiologist. Regions corresponding to cardiac valves and conduction nodes were defined by geometric models. Centroid/bounding box excursion was calculated for cardiac, respiratory, and hysteresis-induced motion. Distance metrics were tested for statistical independence across substructure pairings.

Main Results: 5D-MRI images were successfully acquired and contoured for all volunteers. Cardiac motion was greatest for the coronary arteries (specifically the right coronary) and smallest for the great vessels. Bounding box excursion/HD95 exceeded 1 cm for right heart substructures. Respiratory motion was dominant in the S-I direction and largest for the inferior vena cava. Respiratory hysteresis was generally <5 mm but exceeded 5 mm for some volunteers. For cardiac motion, there were statistical differences between the coronary arteries, chambers, and great vessels, and between the right/left heart. Respiratory motion differed significantly between the base and apex of the heart.

Significance: Our 5D-MRI workflow successfully decouples cardiorespiratory motion with one ~5-minute acquisition. Cardiac motion was >5mm for the coronary arteries and chambers, while respiratory motion was >5mm for all substructures. Statistical considerations and inter-patient variability indicate a subspecific-specific, patient-specific approach may be needed for PRV assessment.

## Introduction

A devasting complication of radiotherapy for lung, breast, esophageal, and other thoracic cancers is cardiotoxicity, which can result in congestive heart failure, coronary artery disease, and pericardial effusion[1–3]. The risk of having a major coronary event increases by 7.4% per additional Gy of mean heart dose during radiation treatments[4], and increased dose to the left coronary artery results in a higher risk of coronary artery disease following treatment[5]. Managing cardiotoxicity in the treatment of lung cancer remains a challenge, with 20% of patients experiencing cardiac issues following treatment[6]. A contributing factor to cardiotoxicity is intrafraction motion, including cardiorespiratory motion, which is difficult to decouple and manage with traditional imaging techniques[7]. Higher doses to specific substructures, namely the coronary arteries[5], cardiac valves[8] and nodes[9], left ventricle[10], and pulmonary vessels[11] have been correlated with specific cardiotoxicities. Thus, a recent thrust has been made to develop techniques for automated cardiac substructure segmentation and to estimate doses to each unique substructure for potential risk assessment and outcome modeling[12–17].

Generally, cardiac motion is not considered when calculating planning organ-at-risk margins (PRVs) for the heart in external beam radiotherapy[18]. However, respiratory motion of the heart is typically managed using gating techniques such as deep inspiration breath holds (DIBH), surface-guided imaging, or tracking respiratory motion of fiducials[19,20]. Current guidance recommends employing a 5 mm threshold for motion management, where motion management techniques should be considered if motion exceeds this threshold in any one direction[21]. Research has shown that cardiac motion differs across cardiac substructures and varies locally, with regions near the base of the heart and specific substructures experiencing displacements greater than 5 mm[22] and even up to 1-2 cm[23]. Despite exceeding the motion management threshold, cardiac motion management is often unachievable in radiation oncology, due to the current lack of cardiac gating capabilities in imaging, treatment planning, and during radiation delivery. Numerous studies have shown that standard PRVs for the heart are insufficient and do not capture the full range of motion of the heart and its substructures, indicating a need for robust, accurate motion quantification for thoracic radiotherapy[23–25].

An emerging cardiac application in radiotherapy that has shown significant promise is the treatment of ventricular tachycardia (VT) with stereotactic body radiation therapy (SBRT)[26,27]. However, cardiotoxicities have been reported following SBRT, as clinical target volume (CTV) margins are typically expanded isotropically to account for motion, exposing unnecessary tissue to radiation[26,27]. Thus, measurement and management of cardiorespiratory motion remains a challenge for VT SBRT procedures. Whether the heart is being spared in thoracic cancer radiotherapy or treated using SBRT for VT, quantifying cardiorespiratory motion and limiting dose to healthy cardiac tissue to reduce cardiotoxicity remain an unmet need.

Cardiac MRI is routinely used in diagnostic imaging to evaluate cardiac anatomy, function, and blood flow[28], and provides sufficient temporal resolution to accurately quantify cardiac motion[23]. The use of cardiac MRI in radioablation of atrial fibrillation and in treatment planning has been explored[29,30], and as MRI simulation becomes increasingly available to clinics, incorporating cardiac MRI into routine radiation oncology workflows becomes possible. The purpose of this work is to implement a novel 5D-MRI workflow[31,32], including reconstructions for both cardiac and respiratory phases from a single MRI acquisition, to derive robust motion models of cardiac substructures to decouple and quantify cardiorespiratory motion. The results of this work may be used to derive accurate PRV margins based on the patients' treatment strategy (e.g., breath-hold or free breathing), to facilitate cardiac-spared treatment planning, or for cardioprotection in targeted VT SBRT.

## Materials and Methods

### 2.1 Image Acquisition and Reconstruction

An overview of our novel 5D-MRI workflow, where data is binned across cardiac and respiratory phases from a single continuous native acquisition, is shown in Figure 1. Data were acquired with a 3D radial balanced Steady-State Free Precession (bSSFP) cardiac MRI sequence with the following parameters: 80,000 radial readouts, echo time/repetition time of 0.9/3.6 ms, flip angle of 35°, imaging volume of $(40.0 \text{ cm})^3$, acquired and reconstructed isotropic resolution of 1.56 x 1.56 x 1.56 mm$^3$, and scan time of ~5 minutes on a 1.5T clinical MRI-simulator (SIGNA Artist, GE Healthcare, Waukesha, WI). Respiratory bellows and pulse oximeter signals acquired during the scan were used for offline retrospective respiratory and cardiac sorting. Data were binned and reconstructed for 10 cardiac phases and 4 respiratory phases (end-inhale (top 10% of the respiratory waveform), end-exhale (bottom 10%), active inhale, and active exhale), where active inhale and active exhale were defined by the remaining 80% of the respiratory waveform between end-exhale and end-inhale. Advanced iterative local low rank reconstruction[33] was used to mitigate undersampling artifacts. The resulting 5D dataset consisted of 40 image volumes for each subject (10 cardiac phases across 4 respiratory phases). A supplemental 3D bSSFP fat-saturated cardiac-gated sequence with navigator-based respiratory gating (3D-Nav) acquired at end-exhale was also included for each volunteer to bolster segmentation ability of the cardiac substructures with the following parameters: echo time/repetition time of 1.8/3.9 ms, flip angle of 70°, imaging volume of (35.0 cm)³-(38.0 cm)³, acquired/reconstruction resolution of 1.70 x 1.70 x 1.80 mm³/0.74 x 0.74 x 0.90 mm³, and scan time of 5 to 9 minutes. 10 healthy volunteers (9M/1F, 23 to 65 years old) were recruited for this study and scanned after providing informed consent.

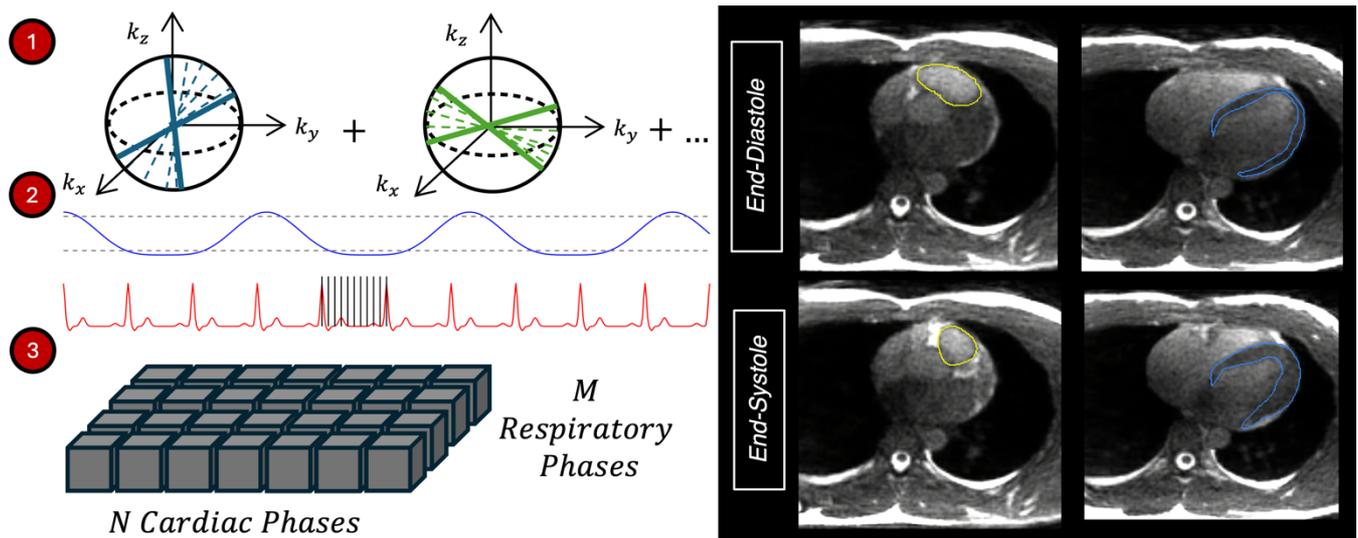

*Figure 1: (Left) Schematic for 5D-MRI workflow where a 3D, radially sampled, bSSFP cardiac MRI is used for acquisition (1), with respiratory bellows and pulse oximeter signals used for retrospective reconstruction (2). Amplitude-based respiratory binning with M phases (dashed lines), and phase-based cardiac binning with N phases (solid lines), yields N x M reconstructed image volumes per dataset (3). (Right) Example 5D-MRI images for end-exhale, end-systole/diastole, where changes in RV shape and position (yellow) and LV blood pool/myocardium (blue) are observed.*

### 2.2 Motion Analysis

#### 2.2.1 Segmentation.
A 15 substructure cardiac model was adopted for motion analysis, including the chambers (left/right ventricle (LV/RV), left/right atrium (LA/RA)), coronary arteries (right (RCA), left main (LMCA), left circumflex (LCX), and left anterior descending (LADA) arteries), ascending/descending aorta (AA/DA), pulmonary veins/artery (PVs/PA), superior/inferior vena cava (SVC/IVC), and whole heart (WH). All substructures were first manually delineated on the end-exhale, end-diastole phase of 5D-MRI using well-established radiotherapy contouring atlases[34,35], using the 3D-Nav as a reference to improve segmentation accuracy when needed. The proximal segments of the RCA, LADA, and LCX were delineated as regions near the base of the heart are most mobile during the cardiac cycle[36] and are most radiosensitive[37]. Contours were then propagated to the remaining 5D-MRI phases using a custom deformable image registration workflow in MIM Maestro (MIM Software, Cleveland, OH), and all propagated contours were visually inspected and modified to match the underlying anatomy. For the coronary arteries, AA/DA,

IVC/SVC, and PA/PVs, contour volume was conserved across 5D-MRI phases to reduce potential delineation errors when truncating substructures within each image. This was not necessary for the chambers and WH as the entire substructure was contoured. For consistent delineation of the proximal coronary artery branches across subjects, definitions from Duane et al.[34] were followed, with all arteries delineated using a 3 mm brush size. For each respiratory phase, contours were delineated for end-systole and end-diastole instead of all 10 cardiac phases as these two phases have been shown to be sufficient for calculating the full extent of cardiac motion[38]. Final contours were verified by a cardiovascular radiologist with 10+ years of experience.

Emerging literature has suggested the importance of cardiac valves and nodes in cardiotoxicity, thus as an exploratory endpoint, a geometric model was adopted for the cardiac valves (mitral valve (V-MV), tricuspid valve (V-TV), aortic valve (V-AV), and pulmonary valve (V-PV)) and conduction nodes (sinoatrial node (SAN) and atrioventricular node (AVN)). Each valve was defined by an 8 mm isotropic expansion of the respective ventricle, followed by a masking (i.e. Boolean intersection) by the corresponding atrium/great vessel following the methods by Finnegan et al.[13] The mitral and aortic valves were defined by an expansion of the left ventricle and masking with the LA and AA, respectively. The tricuspid and pulmonary valves were defined by an expansion of the right ventricle and masking with the right atrium and pulmonary artery, respectively. The SAN and AVN were defined by a sphere with a 1 cm radius, placed at the intersection of the SVC/RA and cardiac chambers, respectively, as defined by Loap et al.[13,39] While these are not direct segmentations, their inclusion in our analysis may provide insights on the mobility of these critical regions of interest.

### 2.2.2 Centroid and Bounding Box Analysis.

The centroid of each contour was found in each image volume and used to calculate centroid displacements across both the respiratory and cardiac cycles for each volunteer, for the anterior-posterior (AP), right-left (RL), and superior-inferior (SI) directions, along with vector centroid displacements ($\sqrt{AP^2 + RL^2 + SI^2}$). Using in-house MATLAB functions (MATLAB 2021b, The MathWorks Inc., Natick, Massachusetts), 3D bounding boxes were automatically calculated for each of the cardiac substructures for each volunteer, throughout all cardiac and respiratory phases, to capture local displacements of each contour. The displacement of the right, left, posterior, anterior, inferior, and superior edges of the bounding boxes were calculated between phases. For both centroid and bounding box displacements, voxel displacements were converted to physical displacements by multiplying by the image resolution. Cardiac motion was defined as motion from end-diastole to end-systole averaged across respiratory states, while respiratory motion was defined as motion from end-exhale to end-inhale averaged across cardiac states. The maximum hysteresis of cardiac substructures due to respiration was calculated following Boldea *et al.* where hysteresis was defined as the distance between pairs of points between exhalation and inhalation trajectories for lung tumors on 4-dimensional computed tomography (4D-CT)[40]. In our implementation, hysteresis was measured by finding the difference in both centroid and bounding box locations between active exhale and active inhale, where non-zero displacement between active exhale and active inhale indicates hysteresis. Displacements due to respiratory hysteresis were averaged across all cardiac phases.

### 2.2.3 Voxelwise Respiratory and Cardiac Displacements.

To describe voxelwise displacements, the distance to agreement between a reference contour "A" and target contour "B" was calculated pointwise (Eqn. 1). Cardiac motion was calculated for each substructure with the reference contour "A" defined at end-exhale, end-diastole and target contour "B" defined at end-exhale, end-systole. Similarly, respiratory motion was calculated with reference contour "A" defined at end-exhale, end-diastole, while target contour "B" was defined at end-inhale, end-diastole.

$$d_a = \min_{b \in B} \|a - b\| \tag{1}$$

From these displacements, the 95% Hausdorff distance (HD95) and mean distance to agreement (MDA) were calculated, defined by the 95th percentile and mean of the displacements, respectively.

### 2.2.4 Statistical Considerations.

Statistical differences in cardiac and respiratory motion were tested for the physiological groupings of substructures (coronary arteries, cardiac chambers, great vessels, and whole heart), individual substructure pairs within each physiological group (e.g. RCA vs LADA for coronary arteries), substructures in the right heart (RCA, RA, RV) vs left heart (LADA, LMCA, LCX, LV, LA), substructures near the base (RCA, LADA, LMCA, LCX, AA, DA, PA, PVs, SVC) vs apex of the heart (LV, RV, LA, RA, IVC), RCA vs right chambers (RA/RV), and LCA vs left chambers (LV/LA), for 8 different distance metrics (centroid displacement in the AP, RL, and SI directions, centroid vector displacements, mean and maximum bounding box displacements, HD95, and MDA). All testing was performed in R using the Kruskal-Wallis nonparametric test with Bonferroni correction, followed by Dunn's test, with a significance level of 0.05 (RStudio v2023.6.1.524, Posit Software PBC, Boston, MA).

## Results

### 3.1 Image Reconstruction and Segmentation

5D-MRI and 3D-Nav images were acquired for all 10 volunteers, and all 15 substructures were successfully contoured on end-diastole/end-systole across all 4 respiratory phases. Figure 2 highlights example 5D-MRI images and delineated chambers/coronary arteries for 4 select cases: subjects with the least, greatest, and average amount of cardiac motion, and a volunteer who had an LADA stent.

### 3.2 Excursion Metrics

Table 1 summarizes the mean AP, RL, SI, and vector centroid displacements for the cohort. Figure 3 highlights the centroid and bounding box excursions due to cardiac and respiratory motion of the volunteer cohort for each substructure. Centroid and bounding box motion of the cardiac valves and nodes are displayed in Table A1/A2. HD95 and MDA for all substructures are shown (Table A3).

### 3.2.1 Cardiac Motion.

Isolated cardiac motion was highly substructure-dependent and was greatest for the coronary arteries and smallest for the great vessels. Across all substructures, cardiac motion was generally in the left, anterior, and inferior directions. Overall, the RCA was the most mobile substructure, with vector centroid motion >5 mm in all subjects. When considering bounding box displacements, the RCA had a displacement >1 cm in the left direction on average, with 9/10 subjects experiencing maximal bounding box displacements >5 mm. The left coronary artery (LCA) was additionally mobile, with 5/10, 3/10, and 6/10 subjects with vector centroid displacements >5 mm for the LADA, LMCA, and LCX, respectively. Maximal

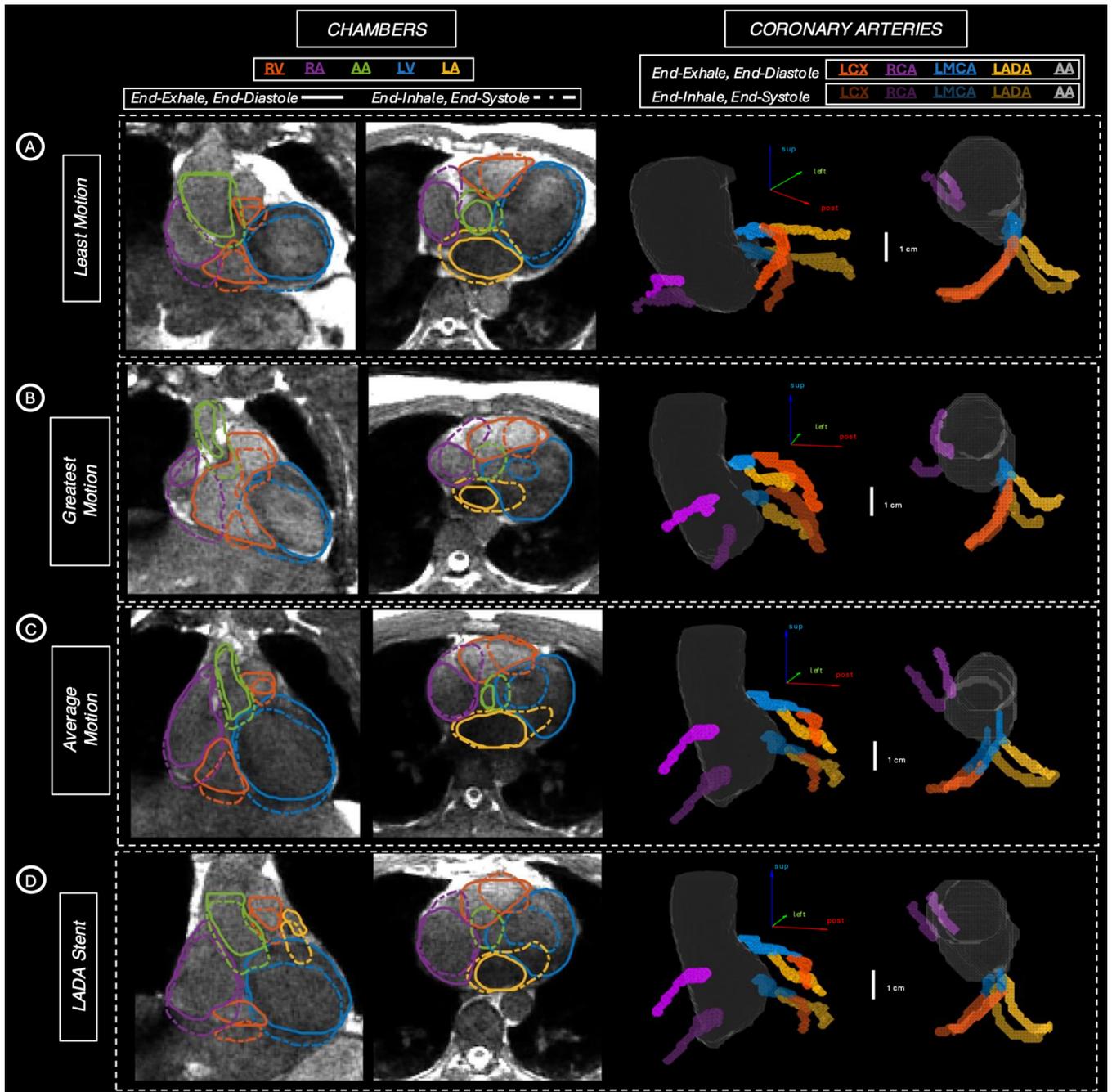

Figure 2: Contours for end-exhale, end-diastole (solid/bright) and end-inhale, end-systole (dashed/dark) on 5D-MRI are shown. Volunteers with the smallest (A), greatest (B), and average (C) magnitudes of cardiac motion, and a volunteer with an LADA coronary stent (D), are highlighted. Substructures generally move inferiorly, anteriorly, and left at end-inhale, end-systole compared to end-exhale, end-diastole. Locally, substructure displacements exceed 1 cm.

bounding box displacements were >5 mm in 8/10, 4/10, and 7/10 subjects for the LADA, LMCA, and LCX. The MDA was >5 mm for the RCA, LADA, and LCX, with HD95 >9 mm for all coronary arteries. Centroid vector displacements were generally <5 mm for the chambers, with RA and RV displacements >5 mm in only 1 subject. However, maximal bounding box displacements were >5 mm in 9/10, 7/10, 5/10, and 9/10 subjects for the LV, RV, LA, and RA, indicating that local displacements are significant for the chambers. While

MDA was <5 mm for all chambers, HD95 exceeded 5 mm for the RA and RV. The great vessels were the least mobile substructures, with vector centroid displacements <5 mm for all subjects and <1-2 mm on average. Maximal bounding box displacements were >5 mm for the AA and PVs in 2/10 and 1/10 subjects, with all other great vessels having displacements <5 mm across all subjects. MDA and HD95 was <5 mm for all great vessels.

**Cardiac and Respiratory Centroid Motion for Volunteer Cohort**

| | | Cardiac Motion | | | | Respiratory Motion | | | |
|---|---|---|---|---|---|---|---|---|---|
| | | R-L (mm) | A-P (mm) | S-I (mm) | Vector (mm) | R-L (mm) | A-P (mm) | S-I (mm) | Vector (mm) |
| Coronary Arteries | RCA | **-5.5 ± 1.7** | 3.8 ± 0.8 | -3.6 ± 2.2 | **8.2 ± 1.2** | 0.5 ± 1.7 | 1.1 ± 1.1 | -4.0 ± 1.5 | 4.6 ± 1.6 |
| | LMCA | -2.3 ± 0.9 | 1.9 ± 1.0 | -3.2 ± 2.2 | 4.8 ± 1.9 | 0.5 ± 1.3 | 0.8 ± 1.1 | -4.5 ± 1.4 | 4.9 ± 1.5 |
| | LADA | -2.0 ± 0.6 | 1.8 ± 1.3 | -4.0 ± 1.8 | **5.3 ± 1.4** | 1.0 ± 2.1 | 1.4 ± 1.3 | -4.9 ± 1.4 | **5.8 ± 1.8** |
| | LCX | -3.0 ± 1.3 | 1.3 ± 1.5 | -3.8 ± 1.9 | **5.5 ± 1.8** | 0.6 ± 0.9 | 0.3 ± 1.1 | -4.6 ± 1.3 | **5.0 ± 1.2** |
| Chambers | LV | 0.0 ± 0.3 | 0.2 ± 0.3 | -0.7 ± 0.4 | 0.9 ± 0.4 | 1.3 ± 1.1 | 1.6 ± 1.0 | **-5.7 ± 1.2** | **6.2 ± 1.4** |
| | RV | -3.4 ± 1.0 | 0.5 ± 0.4 | -1.2 ± 1.0 | 3.8 ± 1.0 | 1.1 ± 1.5 | 1.6 ± 0.5 | -4.3 ± 1.2 | **5.0 ± 1.4** |
| | LA | -1.6 ± 1.0 | 0.8 ± 0.3 | -0.1 ± 0.5 | 1.9 ± 0.9 | 1.1 ± 1.1 | 0.5 ± 0.6 | **-5.1 ± 1.3** | **5.3 ± 1.4** |
| | RA | -1.4 ± 0.9 | 3.4 ± 1.2 | -1.4 ± 1.0 | 4.1 ± 1.4 | 0.1 ± 0.7 | 1.2 ± 0.9 | **-5.2 ± 1.4** | **5.5 ± 1.3** |
| Great Vessels | AA | -0.6 ± 0.6 | 1.3 ± 0.6 | -0.7 ± 0.7 | 1.8 ± 0.8 | 0.7 ± 1.2 | 0.7 ± 0.7 | -3.6 ± 1.4 | 3.9 ± 1.5 |
| | DA | 0.0 ± 0.1 | 0.0 ± 0.3 | 0.3 ± 0.6 | 0.5 ± 0.5 | 0.6 ± 0.5 | 0.9 ± 0.3 | -3.8 ± 1.1 | 3.9 ± 1.2 |
| | PA | -0.2 ± 1.0 | -0.2 ± 0.8 | -0.3 ± 0.2 | 1.3 ± 0.5 | 1.1 ± 2.0 | -0.1 ± 1.2 | -4.0 ± 1.1 | 4.7 ± 1.6 |
| | PVs | 0.2 ± 0.8 | 0.2 ± 0.4 | -0.3 ± 0.2 | 0.7 ± 0.7 | 1.3 ± 1.0 | 0.8 ± 0.6 | **-5.3 ± 1.4** | **5.6 ± 1.5** |
| | IVC | -0.1 ± 0.2 | 0.1 ± 0.2 | 0.0 ± 0.2 | 0.4 ± 0.1 | 0.0 ± 0.7 | 2.3 ± 1.1 | **-5.3 ± 1.8** | **5.9 ± 1.8** |
| | SVC | -0.3 ± 0.3 | 0.5 ± 0.4 | 0.0 ± 0.6 | 0.9 ± 0.5 | 0.0 ± 0.9 | 0.6 ± 0.7 | -4.0 ± 1.3 | 4.2 ± 1.3 |
| Whole Heart | WH | 0.2 ± 0.2 | -0.2 ± 0.2 | 0.4 ± 0.3 | 0.6 ± 0.3 | 1.2 ± 0.9 | 1.3 ± 0.6 | -4.6 ± 0.9 | **5.0 ± 1.0** |

*Table 1: Centroid displacements across 10 subjects for cardiac and respiratory motion in R-L, A-P, S-I directions, and vector displacements. Vector centroid displacement is greatest for the RCA for cardiac motion and the LV for respiratory motion. Motion > 5mm is highlighted.*

**3.2.2 Respiratory Motion.** Isolated respiratory motion varied less across substructures and was dominant in the S-I direction. Isolated respiratory centroid displacements were generally <1-2 mm in the A-P and R-L direction and >4-5 mm in the S-I direction (Table 1). The most mobile substructures due to respiratory motion were the LV and IVC, with 8/10 subjects with a vector centroid displacement >5 mm for both substructures. On the contrary, the least mobile substructures were the SVC and DA, where only 2/10 subjects had a vector centroid displacement >5 mm. More than half of the subjects had vector centroid displacements >5 mm for the LCA, PVs, and IVC. When considering bounding box motion, displacements were generally >5 mm for all substructures, with displacements >5 mm for 8/10 subjects for the RCA, LMCA, AA, and SVC, 9/10 subjects for the LV, RV, PA, and PVs, and 10/10 subjects for the LADA, LCX, LA, RA, DA, and IVC. For the LCA, MDA was >5 mm, while HD95 was >5 mm for all substructures except the DA and WH for respiratory motion.

Motion due to respiratory hysteresis was generally < 5 mm, aside from 3 separate volunteers who exhibited hysteresis-induced motion >5 mm for the RCA, LADA/LCX, and LA/PVs/IVC, respectively. The number of volunteers with cardiac, respiratory, and hysteresis-induced motion exceeding 5 mm for centroid/bounding box motion is shown (Table 2).

**3.2.3 Cardiac Valves and Conduction Nodes.** For the geometric regions representing the conduction nodes and cardiac valves, the V-TV and AVN were the most mobile for isolated cardiac motion, with vector centroid displacements >5 mm for 8/10 and 2/10 subjects. For the V-AV, V-PV, V-

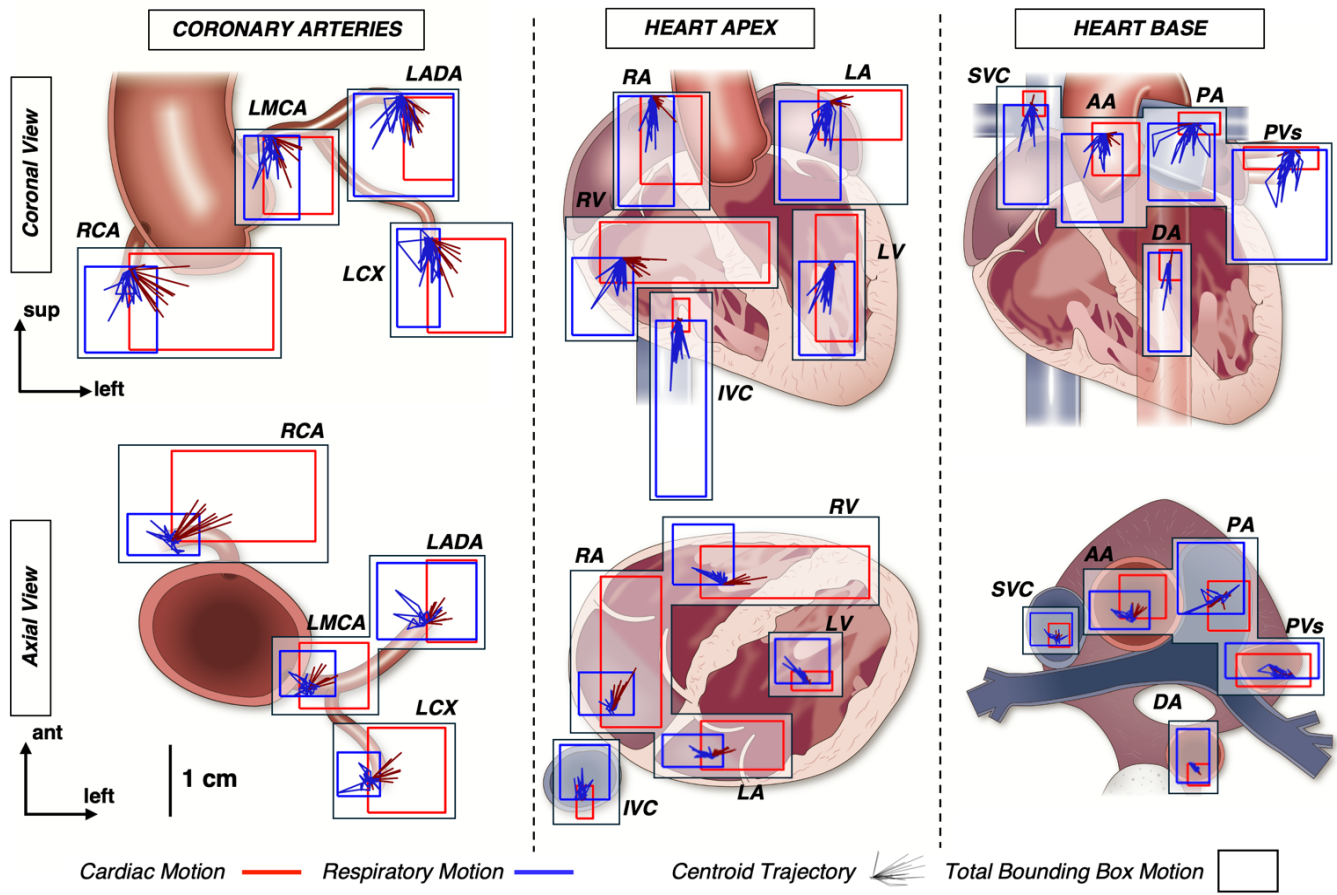

*Figure 3: Cardiac (red) and respiratory (blue) trajectories for coronary arteries (left), heart apex (middle), and heart base (right), for coronal (top) and axial (bottom) views. Centroid trajectories are shown with a solid line for each volunteer, while total bounding box motion for the cohort is shown with a box.*

MV, and SAN, vector displacements were <5 mm for all subjects. Maximum bounding box displacements were >5 mm for the V-AV, V-MV, V-PV, V-TV, and AVN for 4/10, 6/10, 3/10 9/10, and 5/10 subjects, respectively. The SAN was the least mobile, where both centroid vector and bounding box displacements were <5 mm for all subjects. Centroid motion from active-exhale to active-inhale was <5 mm for all substructures across all volunteers, except for the RCA, LCX, and IVC, where 1/10 volunteers had centroid motion >5 mm for each substructure, indicating hysteresis. Similarly, when considering bounding box motion, 1/10 volunteers had excursion >5 mm for the RCA, LADA, LCX, LA, PVs, and IVC. For all other substructures, 0/10 subjects had excursion >5 mm between active-exhale and active-inhale.

Comparatively, vector centroid displacement due to isolated cardiac and respiratory motion for the WH was >5 mm for 0/10 and 6/10 subjects, respectively, with maximum bounding box displacement >5 mm for 1/10 subjects and 8/10 subjects.

### 3.3 Statistical Considerations

A summary of all statistical tests is highlighted in Figure 4, with results of specific metrics given in Figure A1. Across substructure groups, significant differences (P-value < 0.05) in cardiac motion were observed between the coronary arteries (CA)/chambers for 7/8 metrics, between the CA/WH, GA/GV, and chambers/GV for 8/8 metrics, between the chambers/WH for 6/8 metrics, and between the GV/WH for 5/8. Respiratory motion was significant for 4/8 metrics for the chambers/GV, 3/8 metrics for the CA/chambers, and across all pairings for the MDA and HD95. For coronary arteries, the RCA was significantly different from the LCX for cardiac motion for 7/8 metrics and for 8/8 metrics from the LMCA and LADA. All pairings within the left coronary artery were not statistically significant (P-value > 0.05) for all metrics for cardiac and respiratory motion. Respiratory motion was not statistically significant except for the RCA/LMCA for 2/8 and RCA/LCX for 1/8 metrics. For the chambers, statistical differences were observed for cardiac motion for 4/8, 7/8, 4/8, and 5/8 metrics for the LA/RV, LA/RA, LV/RV, and LV/RA, respectively. When comparing the RV/RA and LV/LA, differences in cardiac motion were significant for 2/8 and 3/8 metrics. Across chambers, respiratory motion was not

**Number of Volunteers (out of 10) with Cardiac and Respiratory Motion > 5 mm**

| | | CORONARY ARTERIES | | | | CHAMBERS | | | | GREAT VESSELS | | | | | | WH |
|---|---|---|---|---|---|---|---|---|---|---|---|---|---|---|---|---|
| | | RCA | LMCA | LADA | LCX | LV | RV | LA | RA | AA | DA | PA | PVs | IVC | SVC | WH |
| **CENTROID** | Cardiac Motion | 10 | 3 | 5 | 6 | 0 | 1 | 0 | 1 | 0 | 0 | 0 | 0 | 0 | 0 | 0 |
| | Respiratory Motion | 4 | 5 | 7 | 5 | 8 | 5 | 5 | 5 | 3 | 2 | 4 | 6 | 8 | 2 | 6 |
| | Respiratory Hysteresis | 1* | 0 | 0 | 1** | 0 | 0 | 0 | 0 | 0 | 0 | 0 | 0 | 1*** | 0 | 0 |
| **BOUNDING BOX** | Cardiac Motion | 9 | 4 | 8 | 7 | 9 | 7 | 5 | 9 | 2 | 0 | 0 | 1 | 0 | 0 | 1 |
| | Respiratory Motion | 8 | 8 | 10 | 10 | 9 | 9 | 10 | 10 | 8 | 10 | 9 | 9 | 10 | 8 | 8 |
| | Respiratory Hysteresis | 1* | 0 | 1** | 1** | 0 | 0 | 1*** | 0 | 0 | 0 | 0 | 1*** | 1*** | 0 | 0 |

*Table 2: NunTable 3: Number of volunteers out of 10 with centroid (top) and bounding box (bottom) displacements > 5mm for cardiac, respiratory and hysteresis-inhysteresis motion. Different volunteers exhibiting motion > 5mm for respiratory hysteresis are differentiated with asterisks.*

statistically significant, aside from the LV/RA for 2/8, LV/LA for 3/8, and RV/LA for 1/8 metrics.

For all pairings of great vessels, cardiac motion is statistically significant for <4/8 metrics for all pairings of substructures, aside from the IVC/AA and AA/DA. For the IVC/SVC, respiratory motion was statistically significant for 4/8 metrics, while all other pairings of great vessels were significant for <4/8 metrics for respiratory motion. Comparing all left substructures (LADA, LMCA, LCX, LV, LA) to all right substructures (RCA, RV, RA), differences in cardiac motion and respiratory motion were significant for 6/8 and 3/8 metrics, respectively. When comparing substructures near the base of the heart to substructures near the apex, cardiac motion was statistically different for 0/8 metrics, while respiratory motion was statistically different for 6/8. Comparing the RCA vs right chambers and the LCA vs left chambers, cardiac motion is significant for 8/8 and 7/8 metrics, respectively, while respiratory motion is significant for 2/8 metrics.

## Discussion

This work implemented a novel 5D-MRI workflow to quantify cardiac substructure excursion by decoupling cardiorespiratory motion and providing isolated cardiac/respiratory excursions in a cohort of 10 healthy volunteers. 5D-MRI datasets were acquired in ~5 minutes with sufficient image quality to contour all 15 substructures, including the coronary arteries. Centroid and bounding box displacements, HD95, and MDA were used to describe the full extent of motion for each substructure.

Overall, isolated cardiac motion was dominant in the left, anterior, and inferior directions. In a study by Shechter *et al.,* which used biplane coronary angiography images to quantify coronary artery motion, cardiac motion was similarly found to be dominant in the left, anterior, and inferior directions[36]. Figure 2 highlights that cardiac substructure motion is subject-specific, with some subjects demonstrating

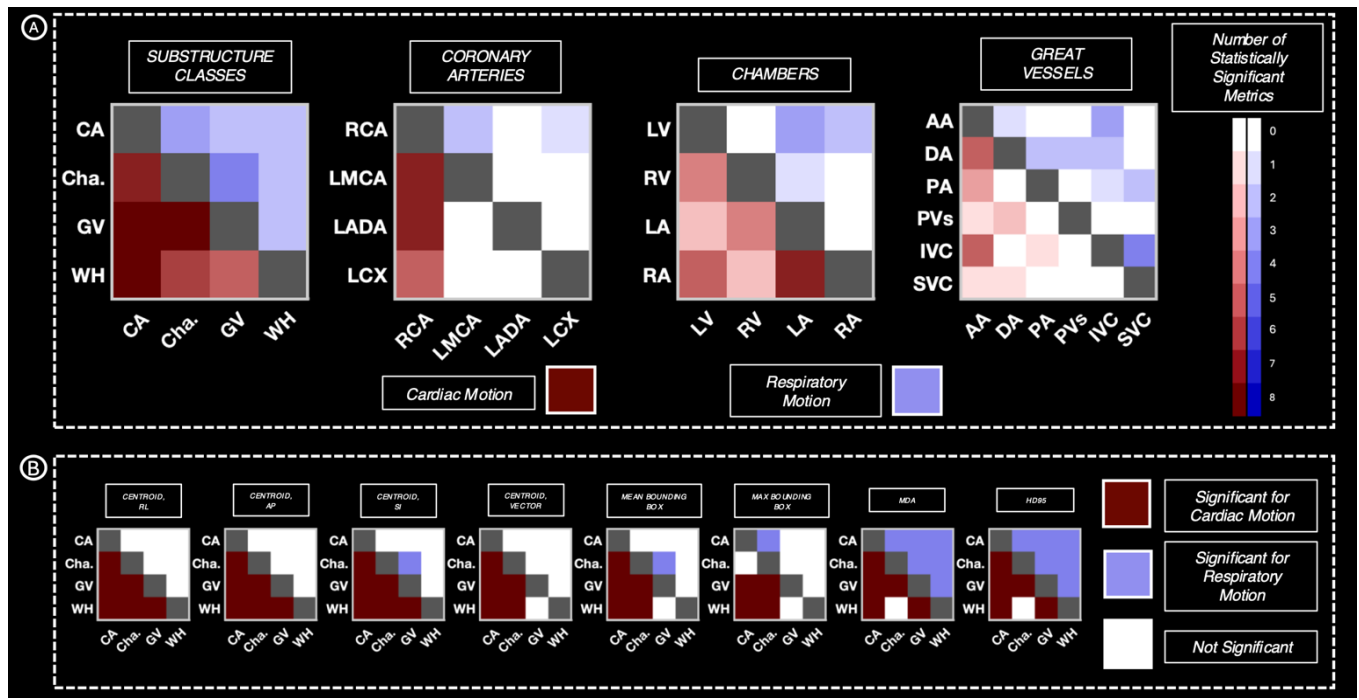

*Figure 4: Summary of statistical significance testing results between all pairs of substructure classes, coronary arteries, chambers, and great vessels, for 8 distance metrics (centroid in RL, AP, and SI directions, centroid vector, mean/max bounding box, MDA, and HD95), with results from all 8 tests shown for substructure classes (B). Number of statistically significant metrics for cardiac motion (red) and respiratory motion (blue) is shown. Detailed results of all statistical tests can be found in Figure A1. Cha. = Chambers.*

RCA, RV, or RA displacements >1 cm, and others experiencing displacements less than 5 mm. Across the 10 volunteers in our cohort, cardiac motion was greatest for the RCA with centroid displacements >5 mm for all volunteers and maximum bounding box displacements greater than 1 cm on average. In the work by Ouyang *et al.*, which used cardiac-gated CT scans from a cohort of 10 patients to measure cardiac motion, centroid displacements were generally <5 mm for cardiac substructures although some patients had RCA, RV, and LADA displacements >5 mm[22]. Greater RCA displacement >1cm compared to the LCA due to cardiac motion have been similarly observed in Tan *et al.*, where cardiac substructure displacement was measured using ECG-gated CT[41]. Johnson *et al.*, which used biplane coronary angiography to quantify the 3D displacement of coronary arteries, found that overall RCA displacement was >2 times larger than the LCA[42]. Statistical differences between the RCA/LCA and right/left chambers were observed for >4/8 metrics, while differences between the LADA/LCX/LMCA, RV/RA, and LV/LA were not statistically significant for >4/8 metrics. Additionally, when the right and left substructures were grouped together and tested for statistical significance, 6/8 metrics were significant for differences in cardiac motion. These results indicate that in our volunteer cohort, cardiac motion was statistically different between the right and left sides of the heart. While there were statistically significant differences between the great vessels for cardiac motion, all great vessels experienced centroid and bounding box displacements <5 mm, indicating that these differences do not meet established criteria for motion management. However, in specific cases where increased sparing of the great vessels is desired, these differences may become important for margin assessment.

Centroid excursion due to respiratory motion was generally >5 mm for cardiac substructures, with average bounding box excursions exceeding 5 mm for all substructures (Table 1). In our volunteer cohort, the LV and IVC were the most mobile structures due to isolated respiratory motion. Other studies have suggested that respiratory motion is greater for the apex compared to the base of the heart[36], which was also observed here. Differences in respiratory motion between the substructures near the base of the heart and substructures near the apex of the heart were found to be statistically significant for 6/8 metrics (Fig. 4), with substructures near the base of the heart (i.e. LV and IVC) experiencing greater motion. Miller *et al.* found that the IVC and RCA experienced the largest respiration-induced displacement, using 4D-CT and T2-weighted MRI to quantify respiratory motion of cardiac substructures[43]. While the RCA was found to have less respiratory motion than the IVC or LV in our cohort, our coronary artery segmentations included only the proximal regions, while Miller *et al.* used the full length of the arteries[43]. Since these segments are closer to the base of the heart than the medial/distal regions of the coronary arteries, it follows that respiratory motion would be smaller. Excursion due to respiratory hysteresis was <5 mm for all substructures on average across subjects, for both centroid and bounding box excursions. However, for some substructures in some volunteers, there were instances where bounding box and centroid excursion exceeded 5 mm (Table A4), indicating that motion due to respiratory hysteresis is person specific.

Regions representing the valves and nodes exhibited varying magnitudes of cardiac and respiratory motion. The AVN and V-TV had average centroid displacements of >5 mm, while maximum bounding box displacements exceeded 5 mm for the AVN, V-TV, and V-MV. These results suggest that the mitral/tricuspid valves and atrioventricular node may require additional consideration during treatment planning, as these substructures are both highly mobile and correlated with radiation-induced valvular heart disease[8] and arrhythmias[9].

The results of this preliminary work suggest that both the cardiac and respiratory motion of cardiac substructures differs both with physiological grouping, as well as location within the heart (right vs left heart, heart base vs apex). While a larger cohort with the inclusion of patients is required before definitive PRV margins can be calculated, our results suggest that certain substructures may require their own PRV margin, as they have statistically different magnitudes of motion exceeding 5 mm. Previously, cardiorespiratory motion quantification has been evaluated using multi-modality approaches, combining 4D-CT for respiratory and cine MRI sequences for cardiac motion[44], or combining multiple cine MRI at different respiratory states to quantify cardiorespiratory motion[45]. However, multi-modality or multi-acquisition approaches result in longer overall scan time to account for multiple acquisitions that require co-registration that may introduce more uncertainty. Our 5D-MRI workflow uses a single, ~5-minute acquisition, where all cardiac and respiratory phases are natively registered to each other, which eliminates potential registration error and reduces scan time. Further, as data are derived from a single acquisition, our technique offers the benefits of no temporal mismatch or co-registration uncertainties.

One limitation of our work includes the small sample size (10 subjects) and general skew toward younger, male healthy volunteers. As such, the observed displacements and statistical relationships may not necessarily hold for cancer patient populations. Nevertheless, the cohort also included two subjects over the age of 60, with one having a coronary stent in their LADA, suggesting applicability of our 5D-MRI workflow in more complex scenarios. One advantage of our 5D-MRI acquisition is that it is acquired over free-breathing conditions in 5 minutes, which would be well-tolerated by cancer patient cohorts without added breath-hold requirements for populations who may be unable to comfortably perform DIBH. Another limitation is that our 5D-MRI iterative reconstruction process is performed entirely offline in a research computing environment, taking ~20-30 hours to reconstruct all 40 cardiac and respiratory phases. For our 5D-MRI workflow to be implemented clinically, reconstruction time could be reduced by reconstructing fewer phases or leveraging GPU acceleration. Another limitation of this feasibility study was that contours were manually delineated, increasing total workflow time. Sources of error associated with contouring include delineation error, inter-observer variability, and potential blurring due to data binning during reconstruction. However, these sources of error were minimized as much as possible by acquiring a high-quality clinical scan as a reference (i.e. 3D-Nav) to improve segmentation accuracy, verification of contour accuracy by a cardiac anatomy expert, and using a small respiratory bin width of 10%. Amplitude-based respiratory binning is an established technique in 4D-CT, where a window width of 10% is often used[46]. Future work can use these initial healthy volunteer datasets to train AI segmentation pipelines as defined in previous work[12,14] to further reduce uncertainties and shorten the time required for contouring.

In summary, cardiorespiratory motion is highly substructure-specific and person-specific. The RCA experienced the greatest amount of cardiac motion, while respiratory motion was greatest for the IVC and LV across our cohort. Substructures can be grouped according to cardiac motion as follows, in descending order of displacement magnitude: RCA, LCA, right chambers, left chambers, and great vessels. Displacements exceed 5 mm in all groups except the great vessels, and differences in cardiac motion are significant (P-value < 0.05) between groups, yet not significant (P-value > 0.05) between substructures within each group (e.g. LADA and LMCA). Similarly, substructures can be grouped by magnitude of respiratory motion, with substructures near the apex of the heart (i.e. chambers and IVC) experiencing greater displacement than substructures near the base (i.e. proximal CAs, AA/DA, PA/PVs, and SVC). Differences in respiratory motion between the base and apex were statistically significant, while respiratory displacement was not significant between substructures within each grouping (e.g. CAs and AA, or IVC and chambers).

Due to the high inter-subject variability observed in this study, a patient-specific, substructure-specific approach to PRV calculation may be necessary for robust planning. However, during treatment planning, incorporating unique margins for each substructures may be unfeasible, as the number of allowed OARs is often limited by the treatment planning system. To reduce the number of additional OARs, a single PRV margin could be applied to each of the above groups, as the differences in motion between substructure within each group are not statistically significant. Future work includes extension of the 5D-MRI workflow for comprehensive motion quantification in an upcoming cardiac-sparing trial for thoracic cancer patients, and in the radioablation of VT.

## Conclusion

Our 5D-MRI workflow has been successfully demonstrated to decouple and quantify cardiorespiratory motion for coronary arteries, chambers, and great vessels via a single 5-minute free-breathing acquisition. Across our volunteer cohort, cardiac motion was highly substructure specific. The RCA, LCA, and chambers exhibited significant cardiac motion (>5 mm). Respiratory motion was dominant in the superior-inferior direction and was >5 mm for all substructures, with substructures near the apex of the heart experiencing greater respiratory motion than the base. For cardiac motion, statistically significant differences were observed between physiological groupings of the substructures, the right/left heart, and between the RCA/LCA and respective right/left chambers. For respiratory motion, statistically significant differences were observed between the apex/base of the heart. Statistical considerations and inter-patient motion variability indicate a need for substructure-specific, patient-specific margins. Future work includes expanding our 5D-MRI workflow to a patient cohort for a comprehensive margin assessment that can be applied for cardiac sparing in thoracic cancer patient cohorts and VT treatments.